\documentclass[12pt]{article}
\usepackage{graphicx}

\def\L54{L_{54}}
\def\E55{E_{55}}
\def\et3{\eta_3}
\def\th1{\theta_{-1}}
\def\r07{r_{0,7}}
\def\x05{x_{0.5}}
\def\et600{\eta_{600}}
\def\et3{\eta_3}

\def\ve{\varepsilon}

\def\cm{\hbox{~cm}}

\usepackage{url}
\usepackage{hyperref}
\def\ve{\varepsilon}
 \newcommand{\aap}{Astronomy and Astrophysics}

 \newcommand{\apj}{Astrophysical Journal}
 \newcommand{\apjl}{Astrophysical Journal Letters}

\def\ve{\varepsilon}

\def\mathnew{\mathsurround=0pt}
\def\simov#1#2{\lower .5pt\vbox{\baselineskip0pt \lineskip-.5pt
       \ialign{$\mathnew#1\hfil##\hfil$\crcr#2\crcr\sim\crcr}}}

\def\beq{\begin{equation}}
\def\enq{\end{equation}}
\def\bea{\begin{eqnarray}}
\def\ena{\end{eqnarray}}

\def\L54{L_{54}}
\def\E55{E_{55}}
\def\et3{\eta_3}
\def\th1{\theta_{-1}}
\def\r07{r_{0,7}}
\def\x05{x_{0.5}}
\def\et600{\eta_{600}}
\def\et3{\eta_3}

\def\ve{\varepsilon}
\def\cm{\hbox{~cm}}

\def \bibpth {/home/veresp/munka/bib/} 

\newcommand\pubnumber{Article 8 in
eConf C1304143}
\newcommand\pubdate{\today}

\def\support{\footnote{Work supported by NASA grant NNX13AH50G
}}
\def\Title#1{\begin{center} {\Large #1 } \end{center}}
\def\Author#1{\begin{center}{ \sc #1} \end{center}}
\def\Address#1{\begin{center}{ \it #1} \end{center}}

\newcommand\pubblock{\rightline{\begin{tabular}{l} \pubnumber\\
         \pubdate  \end{tabular}}}
\newenvironment{Abstract}{\begin{quotation}  }{\end{quotation}}
\newenvironment{Presented}{\begin{quotation} \begin{center} 
             PRESENTED AT\end{center}\bigskip 
      \begin{center}\begin{large}}{\end{large}\end{center} \end{quotation}}
\def\Acknowledgements{\bigskip  \bigskip \begin{center} \begin{large}
             \bf ACKNOWLEDGEMENTS \end{large}\end{center}}

\def\beq{\begin{equation}}
\def\eeq#1{\label{#1}\end{equation}}
\def\eeqn{\end{equation}}

\def\beqa{\begin{eqnarray}}
\def\eeqa#1{\label{#1}\end{eqnarray}}
\def\eeqan{\end{eqnarray}}

\let\bar=\overbar

\def\L{{\cal L}}

\def\Dslash{\not{\hbox{\kern-4pt $D$}}}
\def\dslash{\not{\hbox{\kern-2pt $\del$}}}

\def\msb{{\bar{\ssstyle M \kern -1pt S}}}

\begin{document}
\begin{titlepage}
\pubblock

\vfill
\Title{Gamma-ray burst models with general dynamics 
and fits to Fermi LAT bursts}
\vfill
\Author{ P\'eter Veres$^*$, P\'eter M\'esz\'aros$^*$, Bin-Bin Zhang\support}
\Address{Department of Astronomy and Astrophysics,
Pennsylvania State University, 525 Davey Lab, University Park, PA 16802, USA\\
$^*$ also: Department of 
Physics, and Center for Particle and Gravitational Astrophysics,
Pennsylvania State University
}
\vfill
\begin{Abstract}
We present a dissipative photospheric model for {\it gamma-ray bursts} where
the usual Band peak around MeV energies arises as synchrotron emission from
around the photosphere. We treat the  initial acceleration in a general way and
the GeV emission arises as the interaction of photospheric radiation and the shocked
electrons at the deceleration radius. We show some applications of this model.
\end{Abstract}
\vfill
\begin{Presented}
Huntsville Gamma Ray Burst Symposium\\
14-18 April 2013 – Nashville, Tennessee
\end{Presented}
\vfill
\end{titlepage}
\def\thefootnote{\fnsymbol{footnote}}
\setcounter{footnote}{0}

\section{Introduction}
Observations of GRBs at GeV energies by Fermi LAT have uncovered new properties
which warrant an explanation.  To explain the GeV emission of gamma-ray bursts
observed by LAT, we worked out the details of a dissipative photospheric
scenario in a magnetically dominated jet and its interaction with the shocked
circumstellar material \cite{Veres+12magnetic}. In this scenario, the MeV peak
is given by synchrotron radiation from close to the photosphere, the GeV
emission is given by the interaction of the photospheric photons with the
shocked electrons at the deceleration radius.  In the magnetically dominated
case the dynamics in the acceleration phase is defined by $\Gamma(r) \propto
r^{1/3}$ \cite{Drenkhahn02}.  We have worked out the details of a more general
model, allowing for the $\Gamma(r) \propto r^{\mu}$ type of acceleration which
encompasses both the magnetic and the baryonic model at the two extremes
\cite{Veres+12fit}. We have shown through detailed fitting that the model can
adequately reproduce the Fermi GBM and LAT observations.  
In \cite{Veres+12peak}, we have proven that GRB 110721A, with an unusually high peak energy (\cite{Fermi+12epeak}), can be explained in terms of a dissipative photosphere model, where the dynamics is baryonic or intermediate between magnetic and baryonic.

\section{Dynamics}
\begin{figure}[!htb]
    \includegraphics[width=.8999\columnwidth,angle=0]{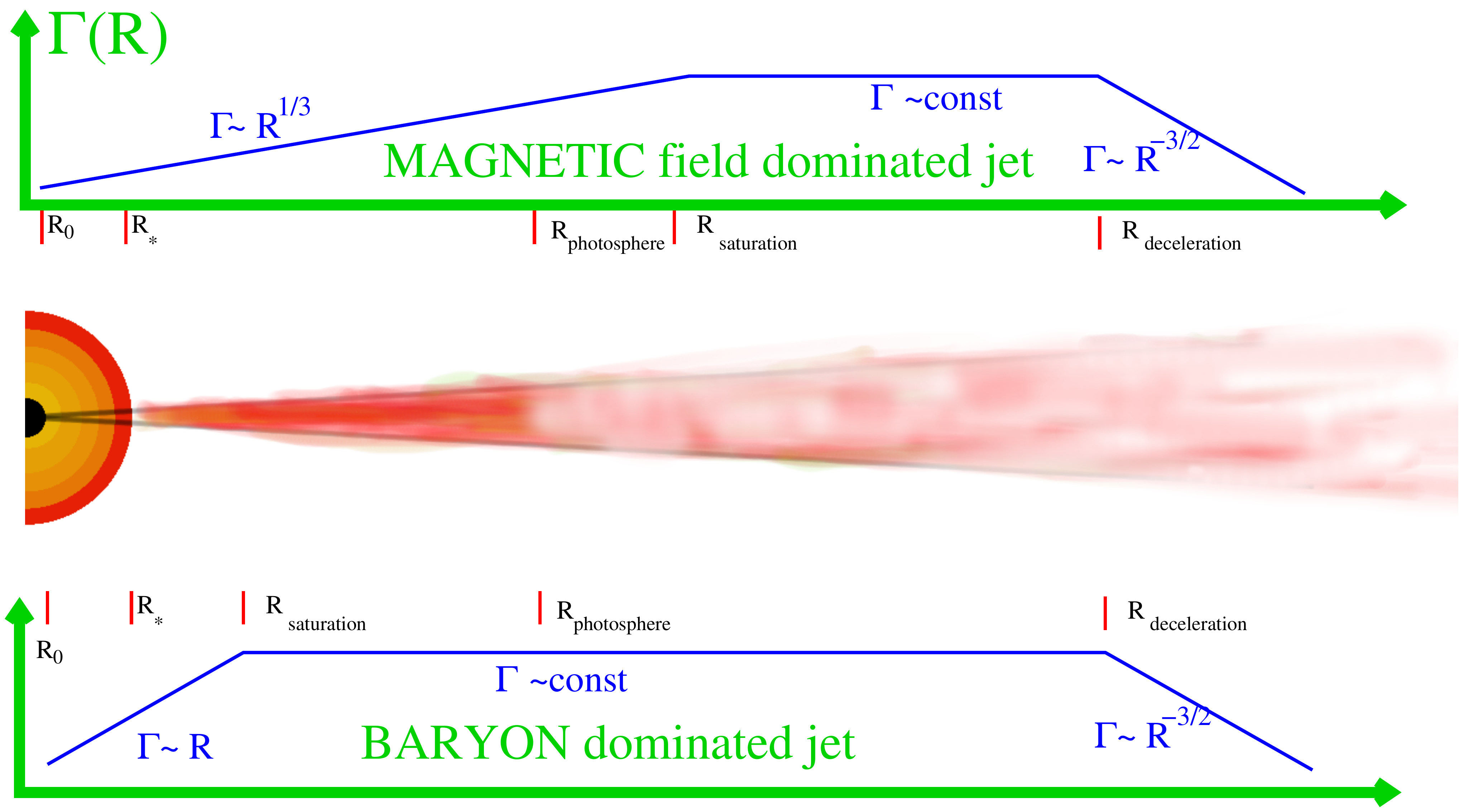}\\
  \caption{Schematic view of the Lorentz factor evolution with radius
  in the magnetically dominated ($\mu=1/3$, top diagram) and in the baryon
  dominated ($\mu=1$, bottom diagram).} 
        \label{fig:cartoon}
      \end{figure}

The jet can accelerate more slowly than in the usual baryonic model (e.g. due
to magnetic dissipation):
\begin{equation}
{\Gamma(r)}
\propto \left\{
\begin{array}{lll}
 r^{\mu}		&	{\rm if}	& r<r_{\rm sat}\\
 {\rm const.}	&	{\rm if}	& r_{\rm sat}<r
\end{array}
\right.
\label{eq:accel}
\end{equation}
We can define a critical Lorentz factor (from $r_{\rm sat}=r_{\rm phot}$),
probing whether the photosphere is in the acceleration or coasting phase of the
dynamics:
\begin{equation}
 \eta_T=\left(\frac{L\sigma_T}{8\pi m_p c^3 r_0}\right)^{\frac{\mu}{1+3\mu}}
\approx \left\{
\begin{array}{lll}
 120~L_{53}^{1/6} r_{0,7}^{-1/6}		&	{\rm if}	& \mu=1/3\\
 1300~L_{53}^{1/4} r_{0,7}^{-1/4}	&	{\rm if}	& \mu=1
\end{array}
\right.
\end{equation}
If $\eta>\eta_T$, the photosphere will be in the  acceleration phase, typical
for {{ magnetically dominated outflows, $\mu\approx 1/3$.}}  In case of
$\eta<\eta_T $, the  photosphere will occur in the acceleration phase, typical
for {{ baryon dominated outflows, $\mu \approx 1$} (see Fig.
\ref{fig:cartoon}).}

The physical parameters with some representative values are $L\approx
10^{53}~{\rm erg/s}$ luminosity, $\eta\approx600$ coasting Lorentz factor,
$r_0\approx 10^7 \cm $ launching radius, $\Gamma_r\approx 1$ shock Lorentz
factor. According to the model, the peak energy dependence is:
\begin{eqnarray}
\ve_{\rm peak} \propto \left\{
\begin{array}{ll}
  L^{\frac{3\mu-1}{4\mu+2}}~ \eta^{-\frac{3\mu-1}{4\mu+2}}~
  r_0^{\frac{-5\mu}{4\mu+2}}~ \Gamma_r^3/(1+z)		&	{\rm if~ }
  \eta>\eta_T	\\
  L^{-1/2}~ \eta^{3}~ \Gamma_r^3/(1+z)		&	{\rm if~ }
  \eta<\eta_T.
\end{array}
\right.
\label{eq:peak}
\end{eqnarray}

A subdominant thermal component is also present:
\begin{eqnarray}
T_{\rm obs} \propto \left\{
\begin{array}{ll}
  L^{\frac{14\mu-5}{12(2\mu+1)}}~ \eta^{\frac{2-2\mu}{6\mu+3}}~
  r_0^{-\frac{10\mu-1}{6(2\mu+1)}} /(1+z)		&	{\rm if~ }
  \eta>\eta_T\\
  L^{-5/12}~ \eta^{8/3}~ r_0^{1/6} /(1+z) 	&	{\rm if~ }
  \eta<\eta_T.
\end{array}
\right.
\end{eqnarray}

\section{Discussion}

	  \subsection{Detailed modeling of the magnetic case ($\mu=1/3$) }
The jet is accelerated by magnetic fields until saturation ($R_{\rm sat}$).
Typically the photosphere occurs above the saturation radius. Semirelativistic
shocks at the photosphere, emit synchrotron radiation and give rise to the Band
peak. External inverse Compton upscattering of the prompt photons on the
shocked electrons at the deceleration radius give the  GeV emission.  Most of
the interaction will occur at $\sim 1/\Gamma$ angle between the photon and
energy momenta.
		 
		 \begin{figure}[htb]
		 \includegraphics[width=.959\columnwidth]{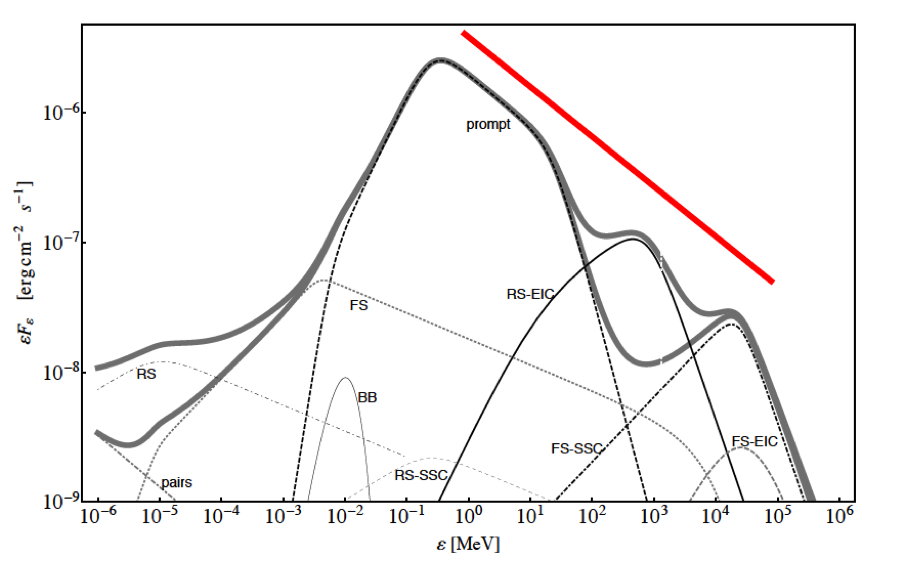}
		 \bigskip\vspace{0.2cm} \caption{Time integrated model spectrum showing
		 various components.  At high energies the components give a simple
		 power law. Upper and lower  thick gray curves represents the spectrum
		 with and without reverse shock. } \end{figure}

	  \subsection{Fitting}
\begin{figure}[!htb]
          \includegraphics[width=.46999\columnwidth,angle=0]{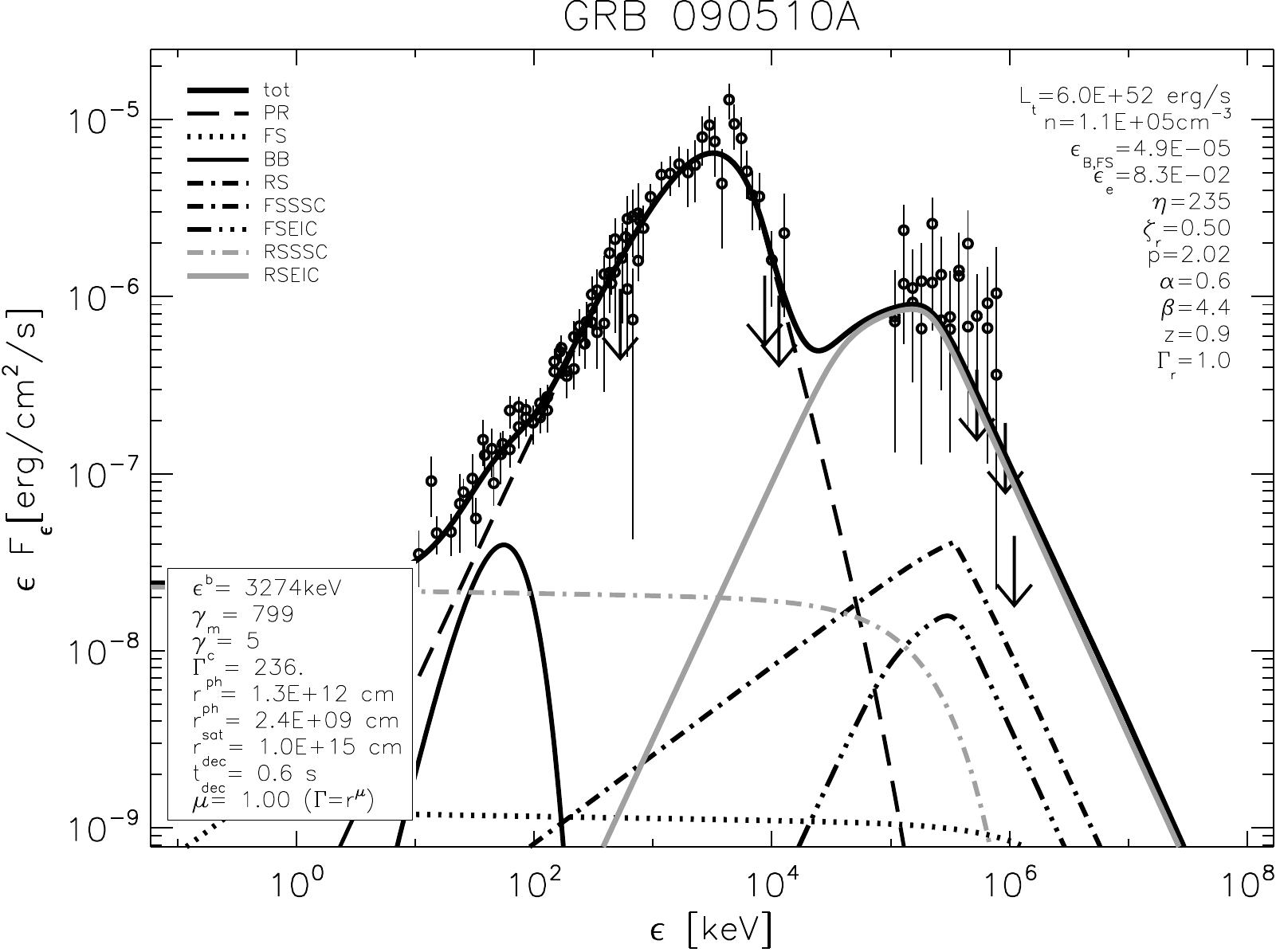}
          \includegraphics[width=.46999\columnwidth,angle=0]{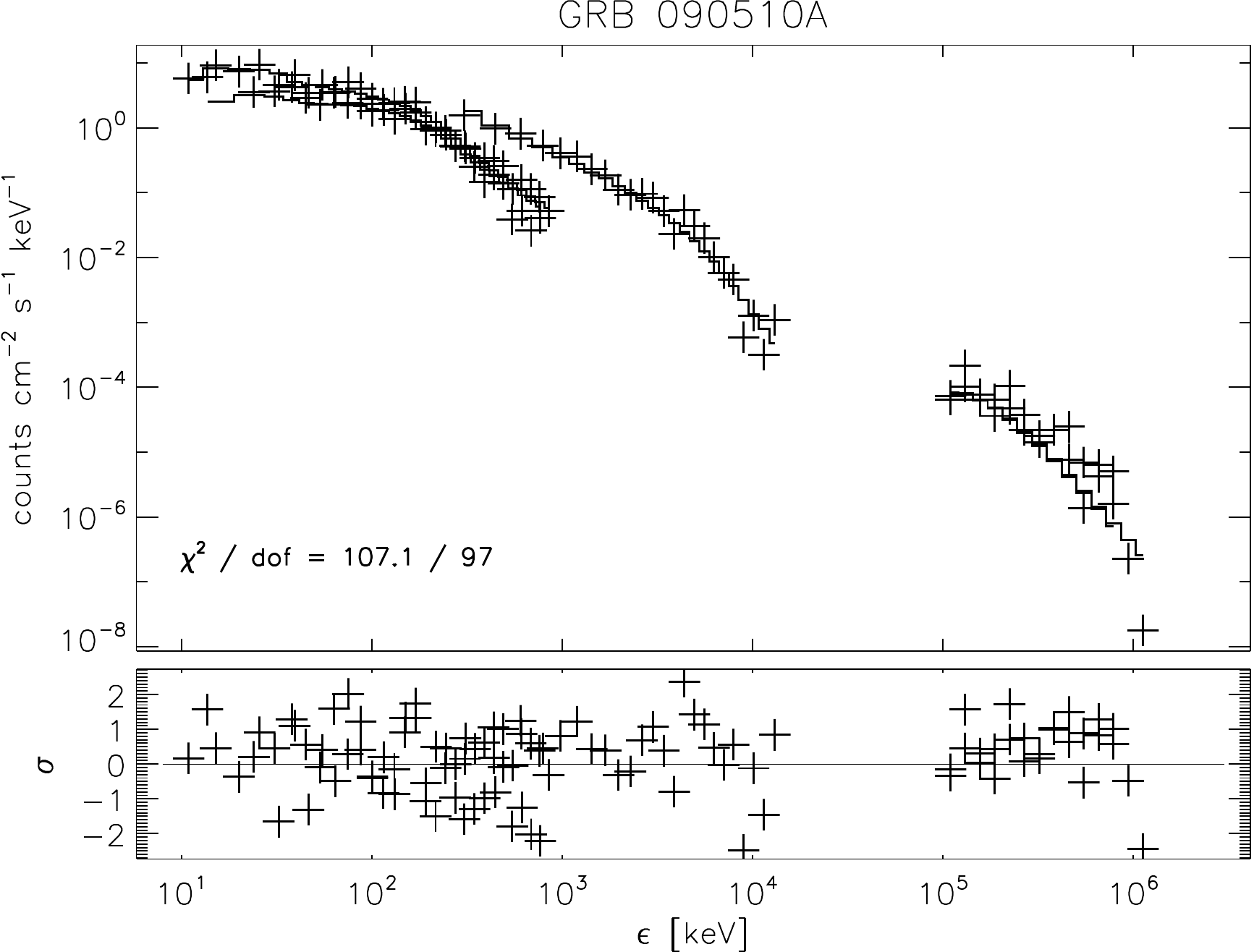}
		  \caption{Example baryonic fit: model spectrum {\it (left)}, count spectrum {\it (right)}. }
		  \label{fig:1}
           \end{figure}

We have developed a model for a general $\mu$ (see eq. \ref{eq:peak}) and used
it to fit four bright Fermi LAT detected GRBs:  080916C, 090510A, 090902B,
090926A \cite{Veres+12fit}.  We use the response matrices of GBM and LAT to
convolve the theoretical spectra and get the count spectrum. We adjust the
parameters of the model to get the best fit.
We show our model with its components on figure \ref{fig:1} as an example for the fit of the baryonic model for GRB 090510A.

	  \subsection{GRB 110721A }
This burst has an {extremely high peak energy ($\approx 15~ $ MeV)} at the
onset. It is hard to reconcile this with the standard internal shock scenario,
but can be accounted for by a dissipative photosphere model with general
dynamics \cite{Veres+12peak}.  Namely, the $\mu=1$ and $\mu=0.5$ cases allow
for the high observed peak energy for reasonable parameters, while $\mu=1/3$ is
hard to reconcile with the observations from this burst.

		 \begin{figure}[!htb]
          \centering
          \includegraphics[width=.49\columnwidth]{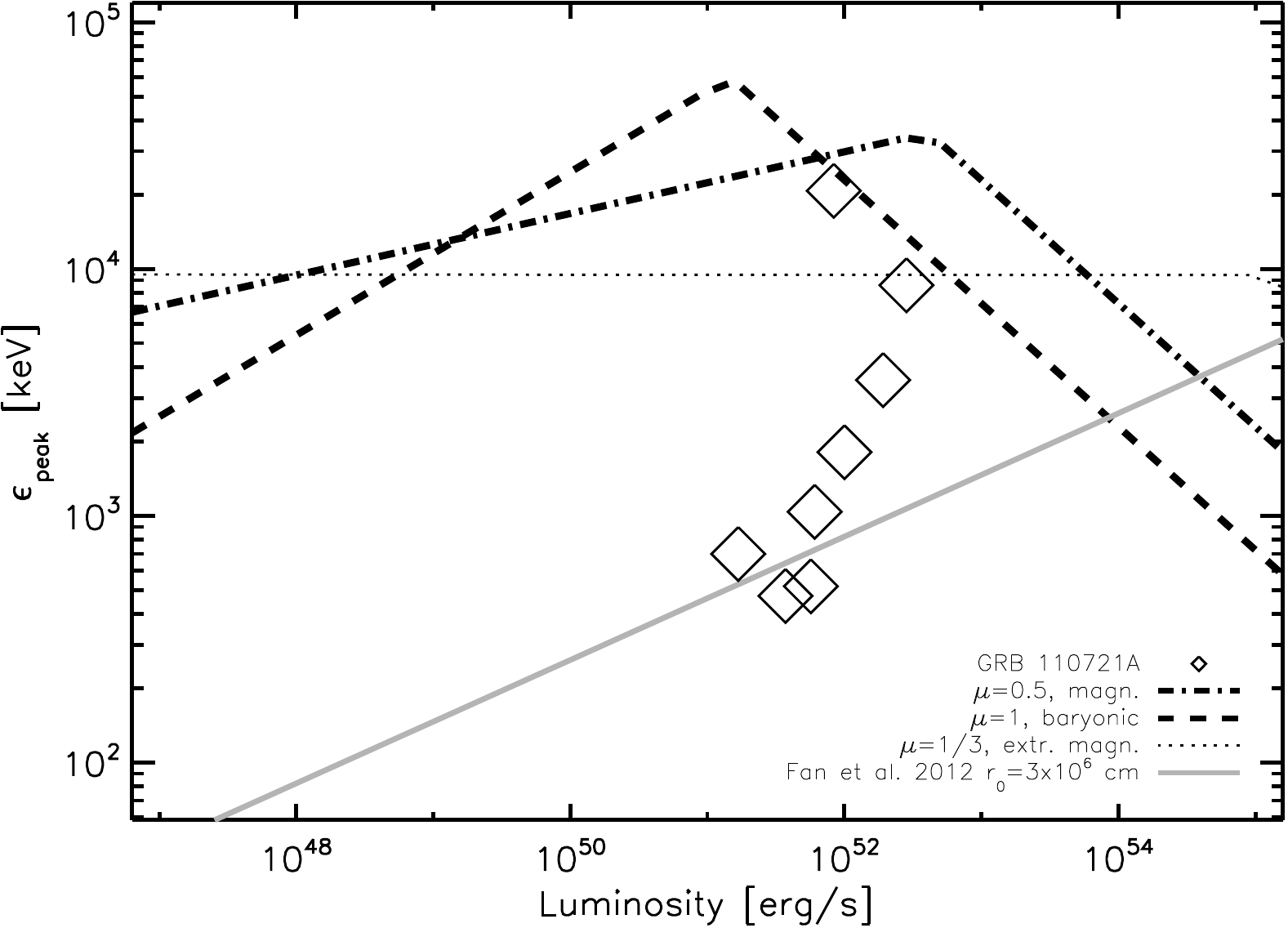}
          \includegraphics[width=.49\columnwidth]{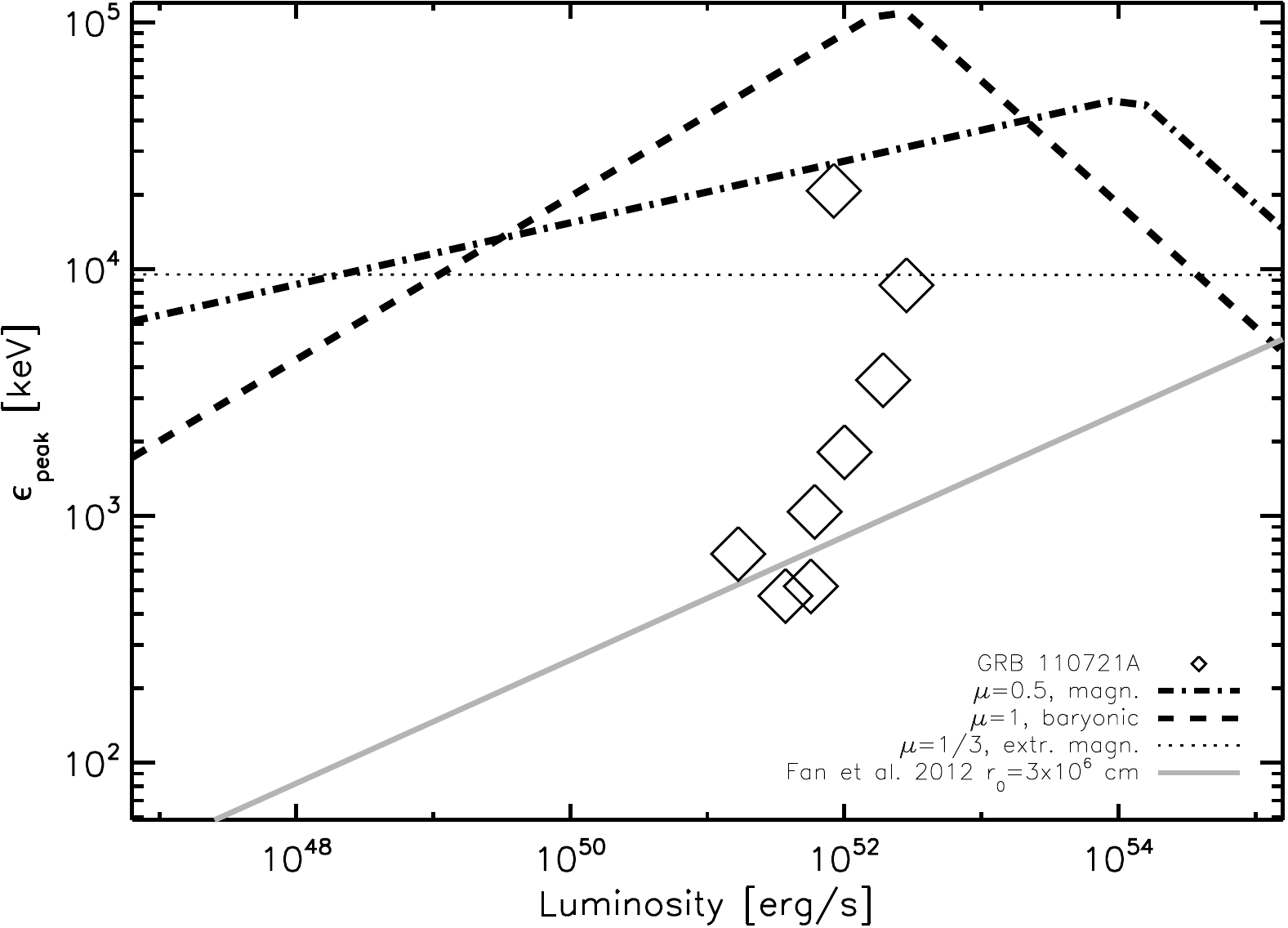}
		  \caption{Peak energy in a dissipative photosphere model with general
		  dynamics. Diamonds show the peak energy and luminosity of GRB 110721A
		  at different times. Dashed, dash-dotted and dotted lines show the
		  limit of the allowed peak energy in the $\mu=\{1, 0.5, 1/3\}$ cases.
		  $\eta=300$ {\it (left)}, $\eta=600$ {\it (right)}}.
           \end{figure}

\Acknowledgements
We acknowledge NASA NNX13AH50G and OTKA K077795 grants (PV, PM), SAO grants GO1-12102X and GO3-14067X (BBZ). We thank Xiaohong Zhao for noticing a typo in Fig. 1.

\bibliographystyle{plain}

\end{document}